\documentclass{IEEEcsmag}

\usepackage[colorlinks,urlcolor=blue,linkcolor=blue,citecolor=blue]{hyperref}
\usepackage{booktabs}
\usepackage{upmath}
\usepackage{graphics}
\usepackage{comment} 
\usepackage{subcaption}
\usepackage{marginnote}
\usepackage{tablefootnote}
\usepackage{multirow}
\usepackage{balance}

\jvol{XX}
\jnum{XX}
\paper{8}
\jmonth{}
\jname{}
\pubyear{}

\usepackage[
backend=biber,
style=alphabetic,
sorting=ynt
]{biblatex}

\title{Bibliography management: \texttt{biblatex} package}
\author{Share\LaTeX}
\date{ }
 
\begin{document}


\title{Sustainable Data Democratization: \\ A Multifaceted Investment for an Equitable Future
\vspace{0.5cm}
}

\author{\IEEEauthorblockN{
Michela Taufer\IEEEauthorrefmark{1}, 
Valerio Pascucci\IEEEauthorrefmark{2},
Christine R. Kirkpatric\IEEEauthorrefmark{3},
and
Ian T. Foster\IEEEauthorrefmark{4}}
\IEEEauthorblockA{U. Tennessee Knoxville\IEEEauthorrefmark{1}, 
University of Utah \IEEEauthorrefmark{2},
University of California San Diego\IEEEauthorrefmark{3},
Argonne National Laboratory \& University of Chicago\IEEEauthorrefmark{4},
\\
Emails: \IEEEauthorrefmark{1}taufer@utk.edu 
}
}

\markboth{Department Head}{Title here}

\begin{abstract}
The urgent need for data democratization in scientific research was the focal point of a panel discussion at SC23 in Denver, Colorado, from November 12 to 17, 2023. This article summarizes the outcomes of that discussion and subsequent conversations. We advocate for strategic investments in financial, human, and technological resources for sustainable data democratization. Emphasizing that data is central to scientific discovery and AI deployment, we highlight barriers such as limited access, inadequate financial incentives for cross-domain collaboration, and a shortage of workforce development initiatives. Our recommendations aim to guide decision-makers in fostering an inclusive research community, breaking down research silos, and developing a skilled workforce to advance scientific discovery.
\end{abstract}

\maketitle


\section{Introduction}
In today's information age, data is the cornerstone of scientific discovery and understanding across diverse domains, from healthcare and education to environmental sciences and technology. 
Data democratization—ensuring that data is accessible and usable for all—presents challenges and opportunities that extend beyond mere technical issues. 
Data democratization should support institutions and individual researchers by providing access to and utilization of both small and large datasets through automated data observatories -- for example, National Aeronautics and Space Administration (NASA) and European Space Agency (ESA) satellites, National Institutes of Health (NIH) funded population studies, and National Oceanic and Atmospheric Administration (NOAA) and Japan Meteorological Agency (JMA) weather and climate monitoring -- that collect large-scale data for community use. 
Most data produced by individuals are never found or accessed, not necessarily because they are not 
FAIR (Findable, Accessible, Interoperable, Reusable)~\cite{Wilkinson2016}, but because they were produced in a way that was useful only for one study. 
The growth in data-driven discovery methods demands increasing focus on “data observatories” in more domains. When dealing with "data observatories," data friction (discrepancies in data formats, heterogeneity of computing systems, and constraints on data sharing imposed by privacy concerns)~\cite{Edwards2013Vast} often poses a significant barrier to data democratization but is far from the only obstacle. 
True democratization of data requires strategic financial, human, and technical investments for an equitable future.

To democratize data successfully, it is essential to adopt a three-pronged strategy based on:
\begin{itemize}
\item {\bf Fostering a collaborative financial approach} that includes industry, academia, and government agencies to develop strategies that acknowledge the costs involved in implementing a solution and balance those costs with the value derived from the associated data.
\item {\bf Promoting equity in training, access, and policies} to enable all segments of society and geographical regions to benefit from scientific advances.
\item {\bf Enhancing AI/ML methods and infrastructure interoperability} to streamline the management and sharing of data across platforms.
\end{itemize}
This article makes the case for the urgency of strategic financial, human, and technological investments needed to achieve sustainable data democratization and advance scientific discovery. 

\section*{FINANCIAL INVESTMENT}


{\it Rapid increase in data production is escalating the financial burdens of data storage and management, posing significant challenges for research institutions and individual investigators.} Data storage and management costs are growing at an alarming rate, posing urgent challenges that threaten the sustainability of scientific progress~\cite{zenodo2024Steinhart, CITI2023}. The cyclical nature of government grants further exacerbates this problem, forcing scientists to rely on uncertain funding streams. This uncertainty poses significant risks of data loss for long-term projects, as escalating recurring storage costs are not matched by steady financial support.

Quantifying the immediate financial costs of data storage and management is straightforward, but assessing the inherent value of data for future scientific discoveries, societal impacts, and knowledge enrichment is more complex. 
This leads to a short-term perspective that penalizes investing in data sustainability. Balancing the tangible, immediate costs against the perceived intangible, deferred benefits puts research institutions in a challenging position, as they must justify immediate expenditures that may not provide direct or guaranteed returns. This dynamic significantly increases the barriers to securing support for long-term data-intensive projects.

Increasingly, the need for big data, for example, to drive new AI-based solutions in fields like environmental science and materials design, requires organized data observatories to generate data at industrial scales. 
In many cases, the costs of establishing and operating these facilities are beyond the scope of public research budgets, which suggests an unappetizing future in which only private corporations have access to vital data assets.  
Public-private partnerships could provide a solution to establishing data observatories in specific fields, such as environmental science. In these partnerships, industry participants pay part of the large costs of data generation in return for privileged access for certain purposes and/or time periods.
Even though there are risks of data privatization with an over-reliance on industry partnerships, dynamic, collaborative ventures between federal agencies, academic institutions, and industry can offer the long-term financial support needed to ensure a continuing data democratization process that may be otherwise unattainable. 

Developing innovative business models that can sustain the financial demands of data management over time is also a necessity.
Models such as subscription services or data-as-a-service offerings can address the challenge of balancing costs with value. 
By generating revenue streams that align with the intrinsic value of data, these models can provide a sustainable financial solution. 
Considering both cost and value in financial strategies is crucial. This integration involves developing policies and practices that recognize and capitalize on the intrinsic value of the data. By doing so, the business models not only cover the costs but also reflect the broader impacts and potential gains of data democratization. This approach ensures a more balanced and equitable consideration of data's worth, aligning financial investments with the long-term benefits that data can offer society beyond individual scientific domains.

\section*{HUMAN INVESTMENT}

{\it Lack of equity in training, access, and policies for data science hinders the development of effective, data-driven solutions to pressing social concerns.} Data-driven tools often embed bias in their inferences when they are designed, tested, and validated by a homogeneous workforce; this process has serious negative consequences for society.
The rapid expansion of complex data has not been matched by a similar expansion of a skilled, diverse workforce that is sensitive to regional and cultural differences.
Conflicts can arise when regional norms significantly differ. 
For example, stringent data privacy regulations in one region may be more relaxed in another, complicating efforts to implement uniform security measures. 
These discrepancies can hinder data sharing and collaboration across borders. 
Harmonious global data ecosystems that are equitable, effective, and ethically sound are contingent upon flexible policies that adapt to various regional contexts. 

Training programs that prioritize equity and actively involve minority-serving institutions (MSIs) and regional institutions are crucial for developing a workforce with the data science skills urgently needed in the scientific community. 
By engaging MSIs, regional institutions, and local stakeholders, these initiatives not only tap into a broader pool of talent but also promote diversity, which is vital for bringing a range of perspectives and innovative solutions to contemporary data challenges. 
Such programs should focus on creating accessible opportunities for all, regardless of socio-economic background, ethnicity, or region, providing tailored resources, mentorship, and support structures to ensure that all participants, especially those from traditionally underrepresented groups, can thrive. Comprehensive training should include foundational data science skills, advanced analytical techniques, and ethical training in data usage. 
Furthermore, these programs should establish strong networks and partnerships among academic institutions, industry leaders, and governmental agencies. 
Using FAIR principles will enable data accessibility and applicability across various domains, institutions, and communities.

Data ecosystems that ensure equitable access to and use of data require policies, security protocols, and privacy-preserving methods that are both universally adoptable and sensitive to local laws and cultural norms. 
This necessity becomes apparent as discrepancies in data privacy laws can disrupt global data-sharing initiatives and collaborative endeavors. 
To this end, communities should develop modular and flexible data management frameworks that not only comply with local regulations but also accommodate differences. 
For example, the CARE Principles for Indigenous Data Governance~\cite{Carroll2020} emphasize the importance of respecting the human aspects of data, advocating for community participation, authority to control, responsibility, and ethics in data management for indigenous communities.
Frameworks ought to incorporate a range of legal and cultural expectations while maintaining essential privacy and security standards. 
Implementing modular policies that can be adjusted according to regional specifics or formulating international accords that set forth fundamental principles and standards for data access and usage tailored to local requirements will enhance widespread data democratization.

\section*{TECHNICAL INVESTMENT}

{\it Opaque and inefficient search, data extraction, and knowledge management solutions impede effective data use.} These solutions struggle to scale due to the increasing volume and complexity of data. Traditional data management methods such as calibration, curation, labeling, and interpretation are often manual, error-prone, and time-consuming, leading to bottlenecks in data accessibility and usability. 
While promising, AI and ML methods for data search, extraction, and management are often opaque to users. The black-box nature of many AI and ML methods results in a lack of trustworthiness of the outcomes. Trustworthiness of results requires transparency at every step of a process, and achieving broad adoption of AI/ML methods requires achieving a level of interpretability and explainability of the results that are currently not attainable. 
Existing infrastructures are not up to the task of efficiently processing and extracting valuable insights from vast arrays of unstructured and structured data.
Furthermore, as datasets grow, determining which data to retain or discard becomes increasingly critical, highlighting the importance of strategic curation and deletion. 
While broadly applicable tools and methodologies have been developed to handle these tasks, execution is frequently domain-specific, demanding an expensive, bespoke approach to each scientific domain.


Despite the acknowledged limitations of current AI and ML technologies in terms of scalability and transparency, they are prioritized due to their exceptional potential for automation and efficiency. As traditional methods often struggle with vast, complex datasets, AI and ML offer adaptive solutions capable of managing them. For example, AI-driven systems automate the categorization and analysis of vast datasets, such as translating 801~TB of satellite images into 54.2~GB of class labels and cloud top and optical properties~\cite{kurihana2022aicca}.
Often used interchangeably in common discourse, in scientific and technical contexts, AI encompasses software and hardware solutions capable of performing tasks that otherwise require human intelligence, including pattern recognition, decision-making, and language understanding over large and complex datasets. ML, a branch of AI, utilizes algorithms to enable computers to analyze data and subsequently learn, predict, or make decisions without explicit programming for each task.
Unlike more resource-demanding, traditional methods, AI and ML have the capacity to swiftly and efficiently process large volumes of data to extract information. 
While the community recognizes that the use of AI and ML technologies is going through growing pains, the community should not only use the current capabilities but also invest in their development to enhance transparency, reliability, and user-friendliness through ongoing research and development.


Integrating flexible and customizable data management systems with intuitive interfaces and advanced visualization tools adaptable to the specific needs of different scientific domains simplifies data management across these fields. 
Adaptable data management systems must be universally applied and flexible enough to meet the unique needs of both research institutions and individual investigators. 
Intuitive interfaces based on powerful tools such as 
Julyter Notebooks~\cite{soton403913} should facilitate easy access to complex datasets for users without deep technical expertise.
Investing in sophisticated data visualization tools such as customizable dashboards publishing FAIR data objects~\cite{10505129} is also critical. These tools transform complex data relationships into transparent, visual formats, facilitating pattern recognition and data analysis. They enhance hypothesis testing and scientific discovery by allowing interactive user experiences with real-time parameter adjustments. 

Integrating data across domains, institutions, and communities is crucial for developing generalizable frameworks that enable efficient data management and analysis. Interoperability and accessibility ensure smooth data exchange and usage across multiple platforms and domains, significantly improving efficiency, fostering collaboration, and sparking innovation across different research domains. Sharing lessons learned and insights from communities with advanced data practices with new users is essential to address disparities in data usage across scientific communities.

\section{CONCLUSIONS}

Sustainable data democratization is not just a technological endeavor but a comprehensive movement that requires concerted financial, human, and technical efforts. By addressing these multifaceted challenges through innovative solutions and collaborative efforts, the community as a whole can pave the way for a future where data is a universal resource, accessible and beneficial to all. This inclusive approach is more than an idealistic goal: it is a necessary step towards a more equitable and informed global society.

\section{ACKNOWLEDGMENT}
This research is supported by the National Science Foundation (NSF) award titled "OAC: Piloting the National Science Data Fabric: A Platform Agnostic Testbed for Democratizing Data Delivery," OAC\#2138811 

\balance

\printbibliography





\end{document}